\documentclass[12pt]{article}
\pdfoutput=1 
\usepackage{graphicx}
\usepackage{authblk}
\usepackage{amsmath}
\usepackage{amssymb}

\newcommand\pubnumber{CIPANP2015-Aguilar-Arevalo}
\newcommand\pubdate{\today}

\def\unam{Universidad Nacional Aut\'onoma de M\'exico, D.F., M\'exico}
\def\umich{University of Michigan, Department of Physics, Ann Arbor, MI, United States}
\def\cabib{Centro At\'omico Bariloche - Instituto Balseiro, CNEA/CONICET, Argentina}
\def\fnal{Fermi National Accelerator Laboratory, Batavia, IL, United States}
\def\kavli{Kavli Institute for Cosmological Physics and The Enrico Fermi Institute, The University of Chicago, Chicago, IL, United States}
\def\ufdrj{Universidade Federal do Rio de Janeiro, Instituto de F\'isica, Rio de Janeiro, Brazil}
\def\zurich{Universit\"at Z\"urich Physik Institut, Zurich, Switzerland}
\def\snolab{SNOLAB, Lively, ON, Canada}
\def\fiuna{Facultad de Ingenier\'ia - Universidad Nacional de Asunci\'on, Paraguay}
\def\niu{Northern Illinois University, DeKalb, IL, United States}

\def\Title#1{\begin{center} {\Large #1 } \end{center}}
\def\Author#1{\begin{center}{ \sc #1} \end{center}}
\def\Address#1{\begin{center}{ \it #1} \end{center}}

\newcommand\pubblock{\rightline{\begin{tabular}{l} \pubnumber\\
         \pubdate  \end{tabular}}}
\newenvironment{Abstract}{\begin{quotation}  }{\end{quotation}}
\newenvironment{Presented}{\begin{quotation} \begin{center} 
             PRESENTED AT\end{center}\bigskip 
      \begin{center}\begin{large}}{\end{large}\end{center} \end{quotation}}
\def\Acknowledgements{\bigskip  \bigskip \begin{center} \begin{large}
             \bf ACKNOWLEDGEMENTS \end{large}\end{center}}

\def\beq{\begin{equation}}
\def\eeq#1{\label{#1}\end{equation}}
\def\eeqn{\end{equation}}

\def\beqa{\begin{eqnarray}}
\def\eeqa#1{\label{#1}\end{eqnarray}}
\def\eeqan{\end{eqnarray}}

\let\bar=\overbar

\def\Dslash{\not{\hbox{\kern-4pt $D$}}}
\def\dslash{\not{\hbox{\kern-2pt $\del$}}}

\def\msb{{\bar{\ssstyle M \kern -1pt S}}}

\begin{document}
\begin{titlepage}
\pubblock

\vfill
\Title{Status of the DAMIC direct dark matter search experiment}
\Author{
  A.~Aguilar-Arevalo$^{a*}$, 
  D.~Amidei$^b$,
  X.~Bertou$^c$,
  D.~Boule$^b$,
  M.~Butner$^{d,j}$,
  G.~Cancelo$^d$,
  A.~Casta\~neda~V\'azquez$^a$,
  A.E.~Chavarr\'ia$^e$,
  J.R.T.~de~Melo~Neto$^f$,
  S.~Dixon$^e$,
  J.C.~D'Olivo$^a$,
  J.~Estrada$^d$,
  G.~Fernandez~Moroni$^d$,
  K.P.~Hern\'andez~Torres$^a$,
  F.~Izraelevitch$^d$,
  A.~Kavner$^b$,
  B.~Kilminster$^g$,
  I.~Lawson$^h$,
  J.~Liao$^g$,
  M.~L\'opez$^i$,
  J.~Molina$^i$,
  G.~Moreno-Granados$^a$,
  J.~Pena$^e$,
  P.~Privitera$^e$,
  Y.~Sarkis$^a$,
  V.~Scarpine$^d$,
  T.~Schwartz$^b$,
  M.~Sofo~Haro$^c$,
  J.~Tiffenberg$^d$,
  D.~Torres Machado$^f$,
  F.~Trillaud$^a$,
  X.~You$^f$, and
  J.~Zhou$^e$}
\Address{
  $^a$\unam,  \\ 
  $^b$\umich  \\
  $^c$\cabib  \\
  $^d$\fnal   \\
  $^e$\kavli  \\
  $^f$\ufdrj  \\
  $^g$\zurich \\
  $^h$\snolab \\
  $^i$\fiuna  \\
  $^j$\niu    \\
  E-mail: {\tt alexis@nucleares.unam.mx} }

\begin{Abstract}
The DAMIC experiment uses fully depleted, high resistivity CCDs to search for dark matter particles. With an energy threshold $\sim$50~eV$_{ee}$, and excellent energy and spatial resolutions, the DAMIC CCDs are well-suited to identify and suppress radioactive backgrounds, having an unrivaled sensitivity to WIMPs with masses $<$6~GeV/$c^2$. Early results motivated the construction of a 100~g detector,  DAMIC100, currently being installed at SNOLAB. This contribution discusses the installation progress, new calibration efforts near the threshold, a preliminary result with 2014 data,  and the prospects for physics results after one year of data taking.
\end{Abstract}
\vfill
\begin{Presented}
Twelfth Conference on the Intersections of Particle and Nuclear 
Physics (CIPANP 2015)\\ \bigskip
Vail, CO, United States,  19 -- 24 May, 2015
\end{Presented}
\vfill
\end{titlepage}
\def\thefootnote{\fnsymbol{footnote}}
\setcounter{footnote}{0}

\section{Introduction}

A variety of astrophysical and cosmological observations provide strong evidence supporting the existence of non-baryonic, cold dark matter \cite{Clowe:2006,Hinshaw:2013,Planck:2015}. Theoretical models proposing weakly-interacting massive particles (WIMP) with masses in the range of 1-15 GeV/$c^2$ \cite{cohen:2013} have gained interest motivated by recent experimental results \cite{cdmssi:2013,dama:2013}.

The goal of the DAMIC (Dark Matter in CCDs) experiment is to use high-resistivity charge-coupled Devices (CCDs) as detectors for the direct search for light dark matter particles. The low threshold and low background of CCDs, and the relatively small mass of the Si nucleus will allow the experiment to explore new regions of the WIMP parameter space.

The potential of CCDs to perform a WIMP search was succesfully demonstrated in 2011 during a short test run at a shallow depth \cite{barreto:2012}. This promising result prompted the DAMIC collaboration to install a larger detector in the SNOLAB laboratory in Canada. The collaboration is well on its way to installing a 100~g version of the experiment, DAMIC100, which should begin operations during the first half of 2016.

\section{CCDs as particle detectors}

The DAMIC CCDs have 8 to 16 million pixels, with sizes of 15~$\mu$m$~\times~$15~$\mu$m, and thicknesses of several hundred $\mu$m. They have a three-phase polysilicon gate structure with a buried p-channel, and an active region made of high resistivity n-type silicon, whose low donor density ($\sim10^{11}$cm$^{-3}$) allows for fully depleted operation at low bias voltages (40~V for a 675~$\mu$m-thick CCD). Holes produced by ionization (3.62~eV per $e^-$ and hole) difuse as they drift towards the gates attaining a lateral spread that can be used to reconstruct the $z$-coordinate of point-like interactions \cite{chavarria:2015}.

\begin{figure}[t]
\centering
\scalebox{0.27}{\includegraphics{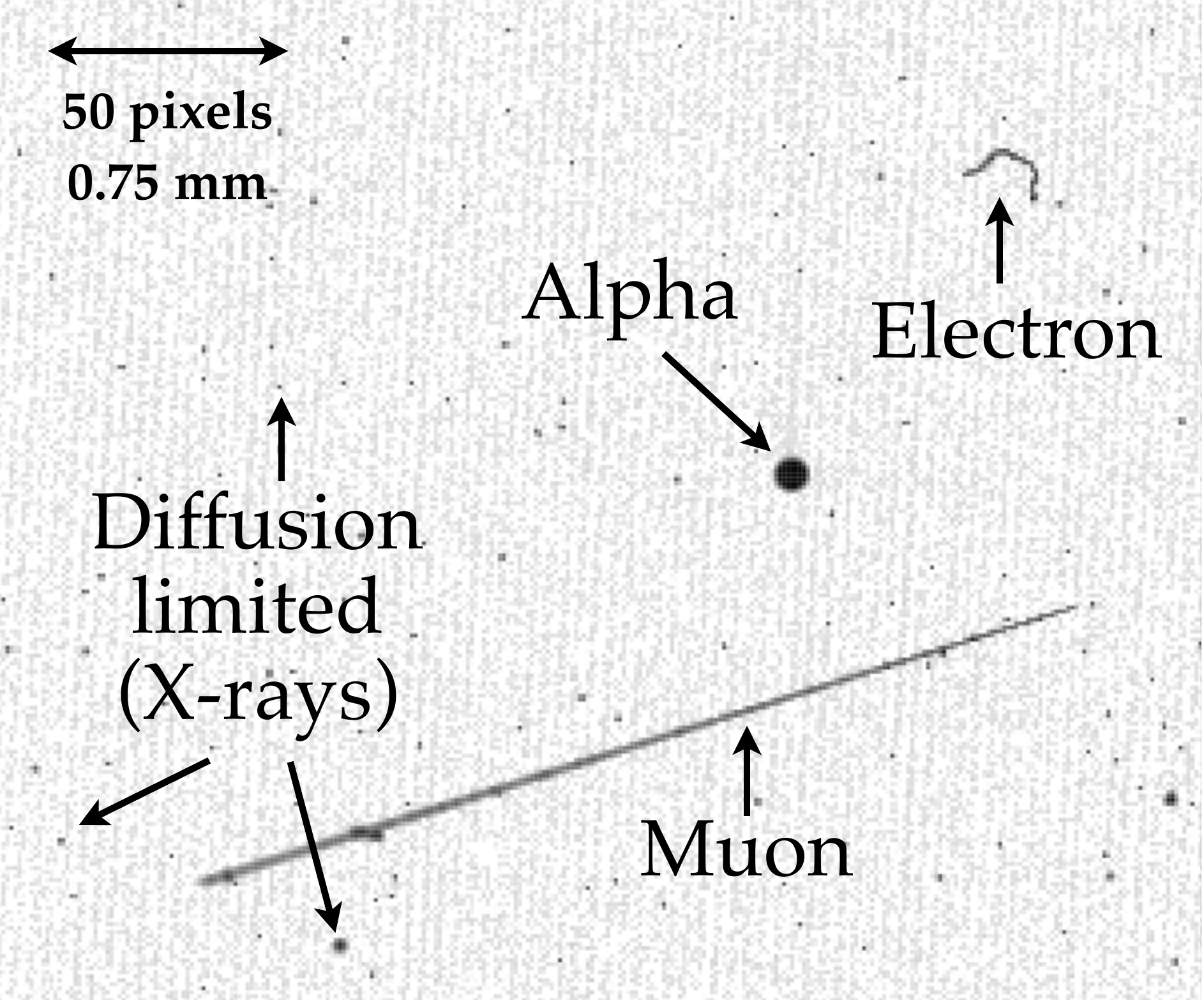}}
\hspace{0.2cm}
\scalebox{0.50}{\includegraphics{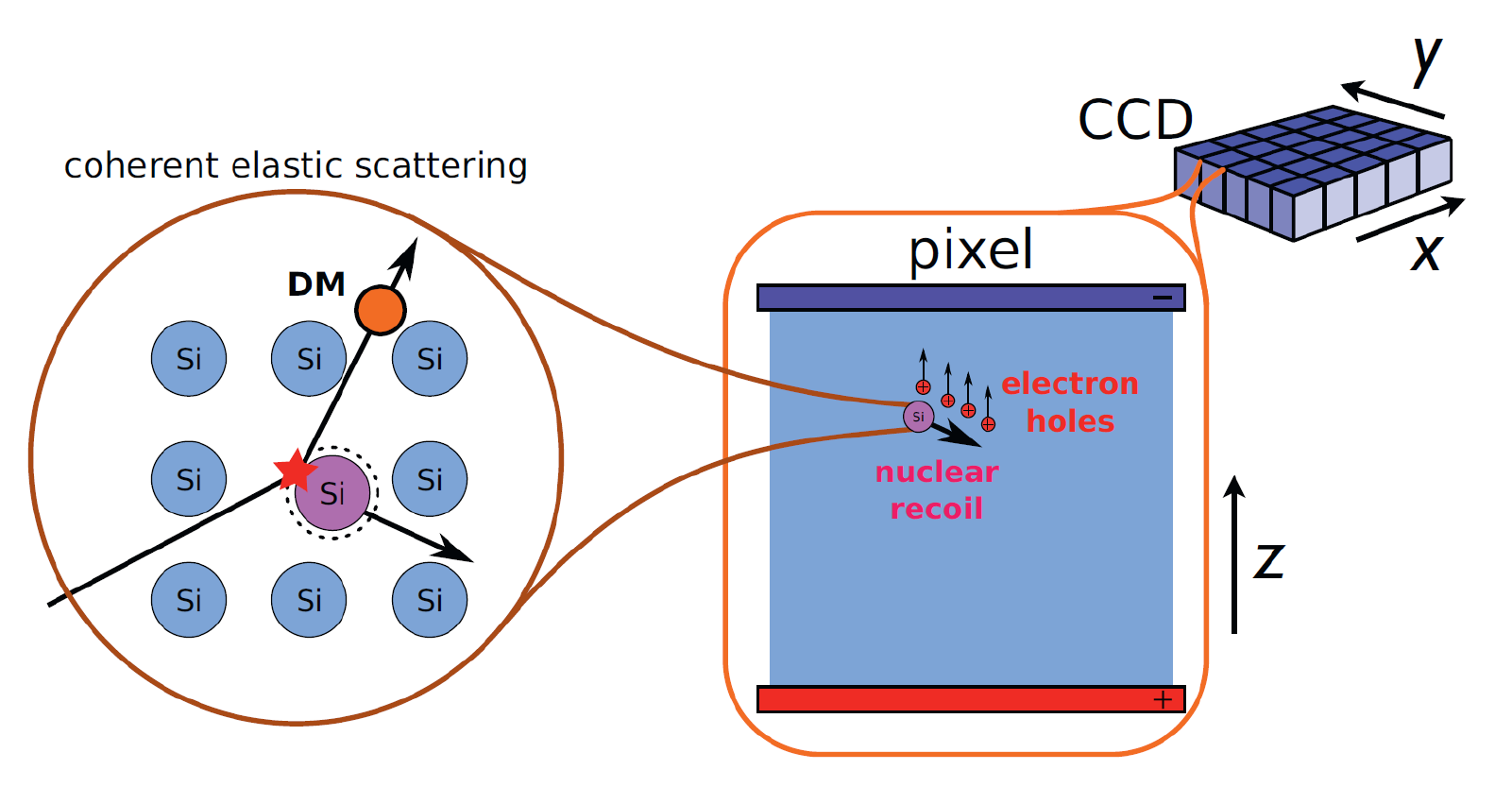}}
\caption{\small
Left: Particle tracks in a DAMIC CCD.
Right: WIMP detection principle.
}
\label{fig:particle-tracks}
\end{figure}

Charge collected over multiple hour exposures ($\sim 8$ hr) is read out row by row through a serial register on one side of the CCD. The pixel charges are measured in a low capacitance output node with a typical RMS noise of $\sim 2~e^-$/pix (or $\sim7$~eV/pix). At the operating temperature of 100~K the dark rate is $<0.01~e^-$pix$^{-1}$day$^{-1}$. The nominal DAMIC pixel threshold of $\sim$50~eV$_{ee}$ is defined by the condition that a pixel value be  7$\sigma$ above the noise. Figure \ref{fig:particle-tracks} shows various particle tracks in a DAMIC CCD (left), and illustrates the WIMP detection principle (right).

Energy calibrations from exposures to X-rays from a $^{55}$Fe source and light-element fluorescence, and $\alpha$ particles from a $^{241}$Am source, have demonstrated the excellent linearity and energy resolution of the DAMIC CCDs \cite{chavarria:2015,tiffenberg:2014}.
Alpha spectroscopy and studies of spatial $\beta-\beta$ coincidences have been used to set stringent constraints on the radioactive contamination in the CCDs \cite{aguilar-arevalo:2015}.
DAMIC's threshold to nuclear recoils corresponds to roughly $\sim$0.5~keV$_{r}$. This estimation has large uncertainty as it is based on the extrapolation of the Lindhard model \cite{ziegler:1985} of the ionization efficiency of nuclear recoils, for which measurements exist only down to 3-4~keV$_r$ \cite{dougherty:1992, gerbier:1990}.

\section{Progress towards DAMIC100}

\begin{figure}[t]
\scalebox{0.48}{\includegraphics{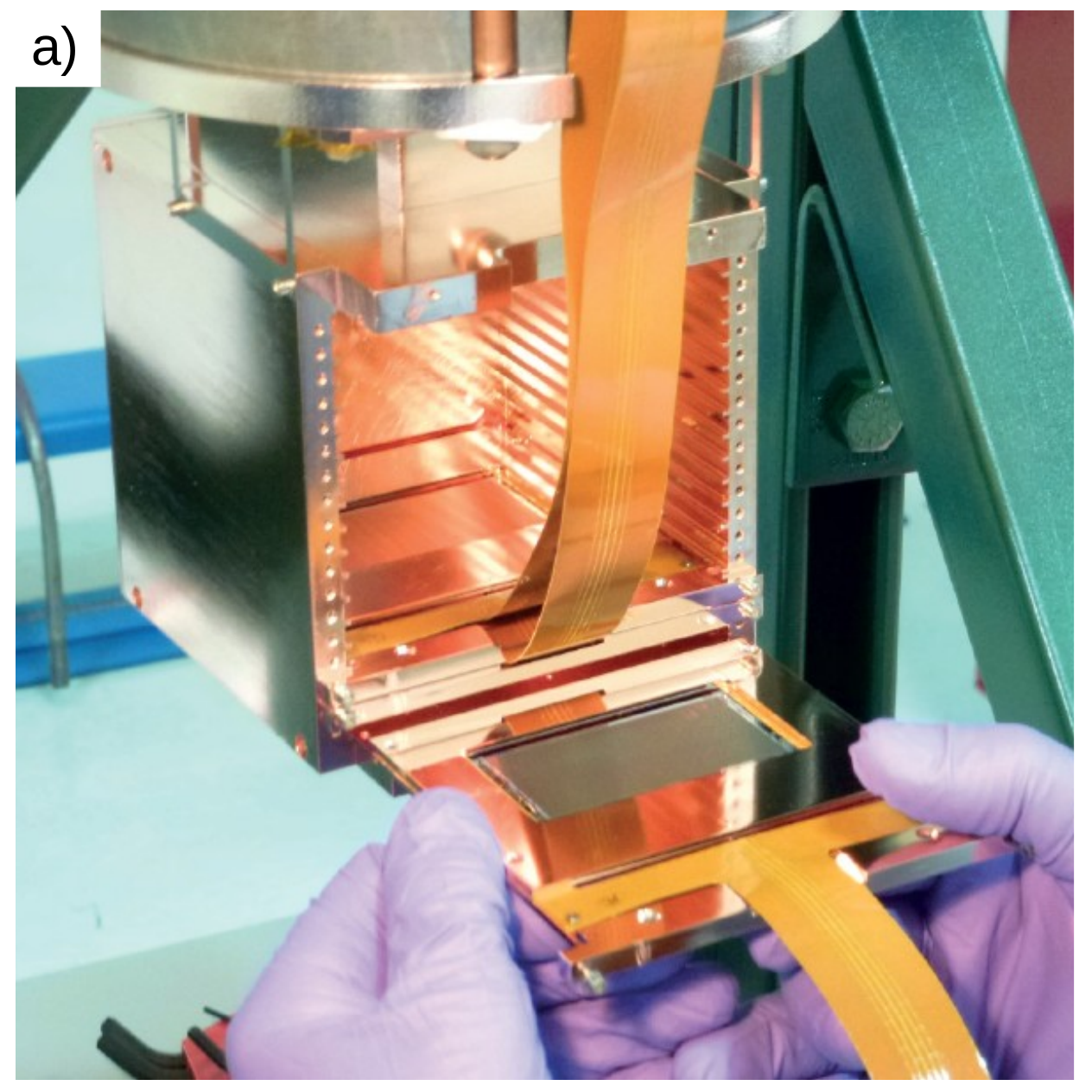}}
\scalebox{0.76}{\includegraphics{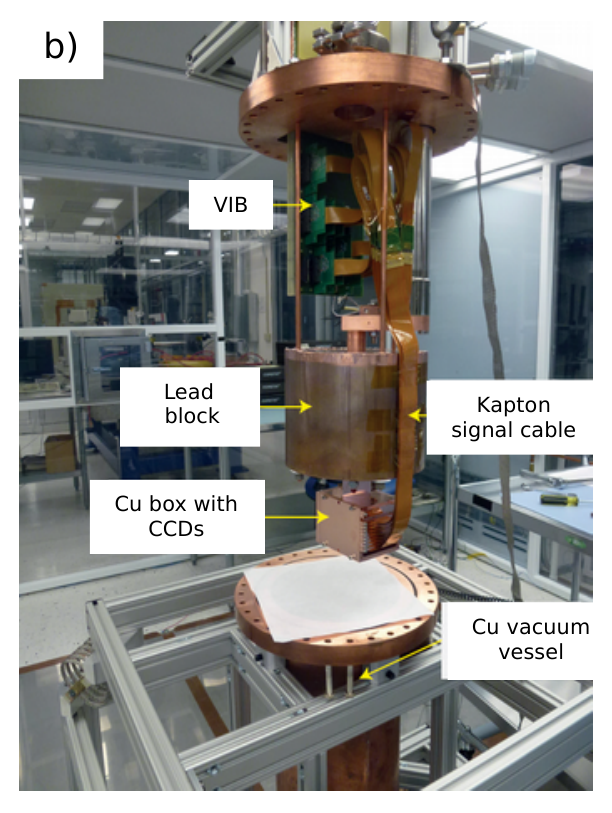}}
\scalebox{0.38}{\includegraphics{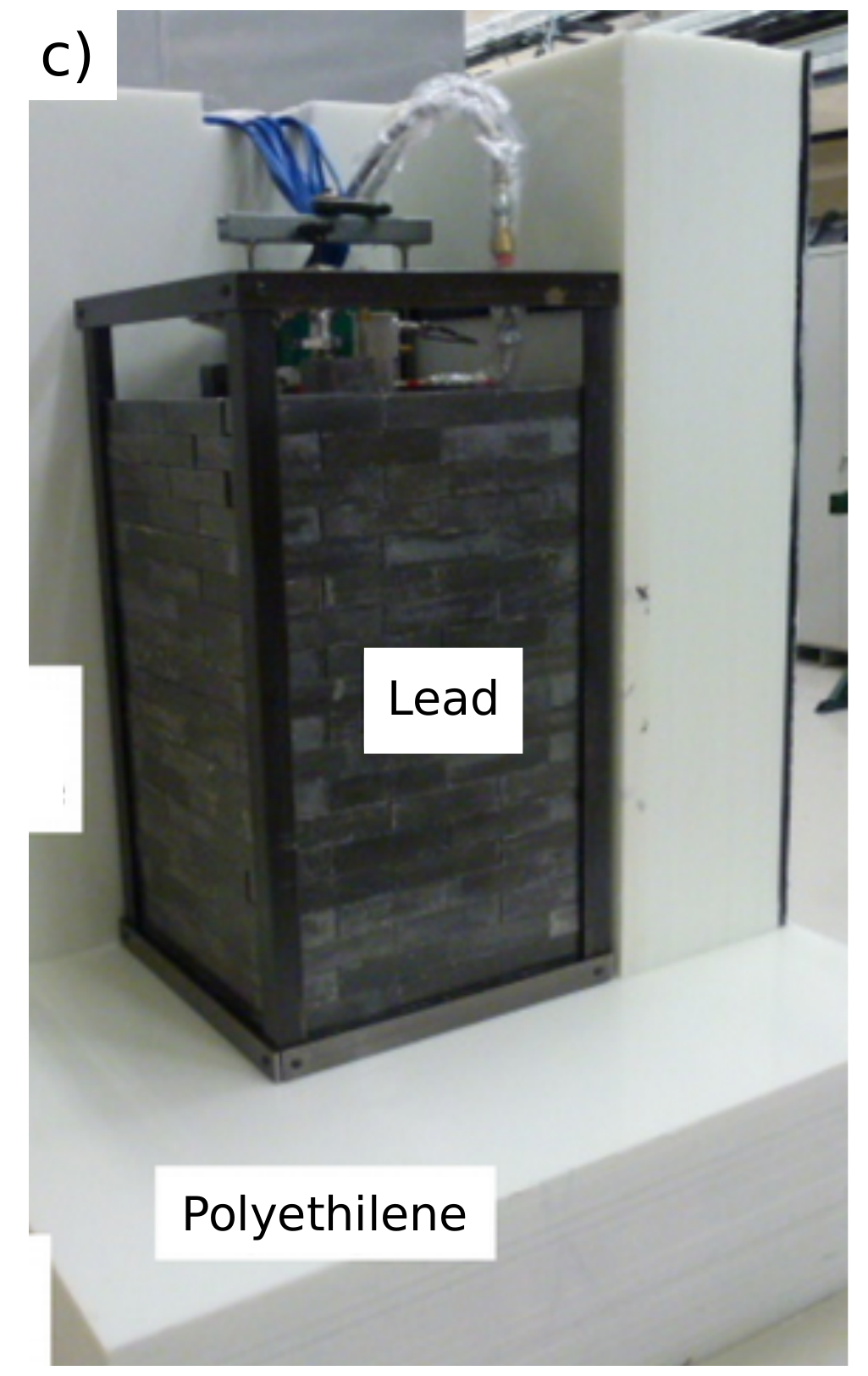}}
\caption{\small
Infrastructure installed at SNOLAB.
a)\,The DAMIC100 copper box with three 2k$\times$4k (8~Mpix) CCDs in the new package.
b)\,Components inside the vacuum vessel.
c)\,Vessel inside the lead castle during installation of the poliethylene shield.
}
\label{fig:initial-setup}
\end{figure}

DAMIC was installed at SNOLAB in November 2012. Details of the initial installation can be found elsewhere \cite{chavarria:2015,tiffenberg:2014}. 
 Each CCD is epoxied to a high-purity Si support piece fitted to a copper bar, which facilitates the handling of the packaged CCD and its insertion into a slot of an electropolished copper box cooled to 100~K inside a copper vacuum vessel ($10^{-6}$~torr), see figure \ref{fig:initial-setup}. The CCDs are connected to a vacuum interface board (VIB) through Kapton flex cables running along the side of an 18-cm-thick lead block hanging from the vessel flange, which shields the CCDs from radiation produced at the VIB. The cables are glued to the support piece and covered with copper frames. The vacuum vessel sits inside a 21~cm-thick lead castle to shield from ambient $\gamma$-rays. The innermost inch of lead comes from am ancient Spanish galleon and has negligible content of $^{210}$Pb, strongly suppressing the background from bremsstrahlung $\gamma$s produced by $^{210}$Bi decays. A 42~cm-thick polyethylene shield is used to moderate and absorb environmental neutrons.

In December 2014, preparations for the deployment of DAMIC100 were started. A major modification was the installation of a new copper box with improved grounding, as shown in figure~\ref{fig:initial-setup}(a), suited to house up to eighteen 4k$\times$4k (16~Mpix), 5.8~g, 675~$\mu$m-thick CCDs with a new package design. At that time, three 675~$\mu$m 2k$\times$4k (8~Mpix) CCDs were installed, pending packaging and testing of the 16~Mpix versions. The background event rate level with this setup was measured to be $\sim50$~dru (1 dru = 1 kg$^{-1}$day$^{-1}$keV$_{ee}$$^{-1}$), two orders of magnitude lower than in the original setup.

A series of R\&D efforts aimed at identifying the source of the observed limiting background have been carried out: a pressurized N$_2$ box was installed in order to reduce the radon in the setup. The outer copper vessel was chemically etched to remove any contaminants that may have accrued since the time of the installation. Radioactivity tests performed by placing ancient lead and copper shields with different surface treatments around the CCDs demonstrated problems with the electropolished copper\footnote{At the time of writing this report, the collaboration has pinned down the source of background to  impurities embeded in the copper used for the box and package holders, possibly during the elecropolishing process.}. Wafers with a total of twenty-four 16~Mpix CCDs to be used for DAMIC100 have been acquired and are in the process of being packaged and tested. At the time of this writing one 4k$\times$4k 16~Mpix CCD with mass of 5.9~g has already been installed at SNOLAB and preliminary tests have shown a background count rate of $<5$ ~dru, approaching the DAMIC100 goal.

\section{Near-threshold response to $\gamma$-rays and neutrons}

\begin{figure}[t]
\scalebox{0.64}{\includegraphics{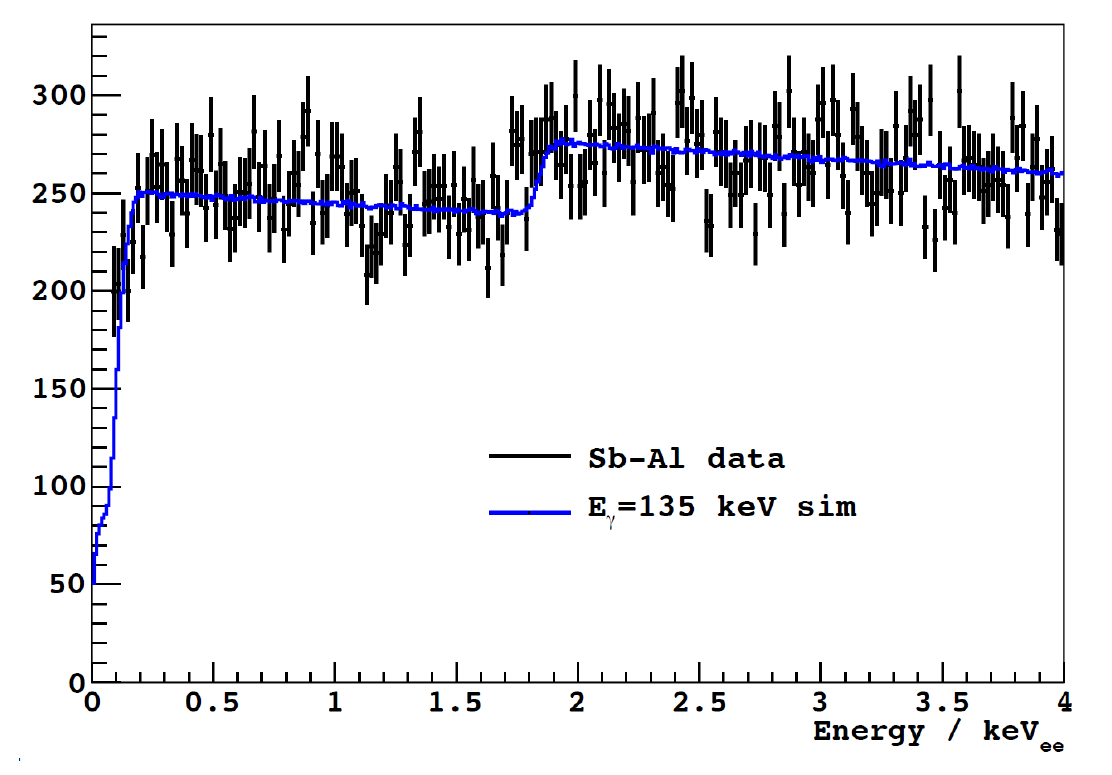}}
\scalebox{0.39}{\includegraphics{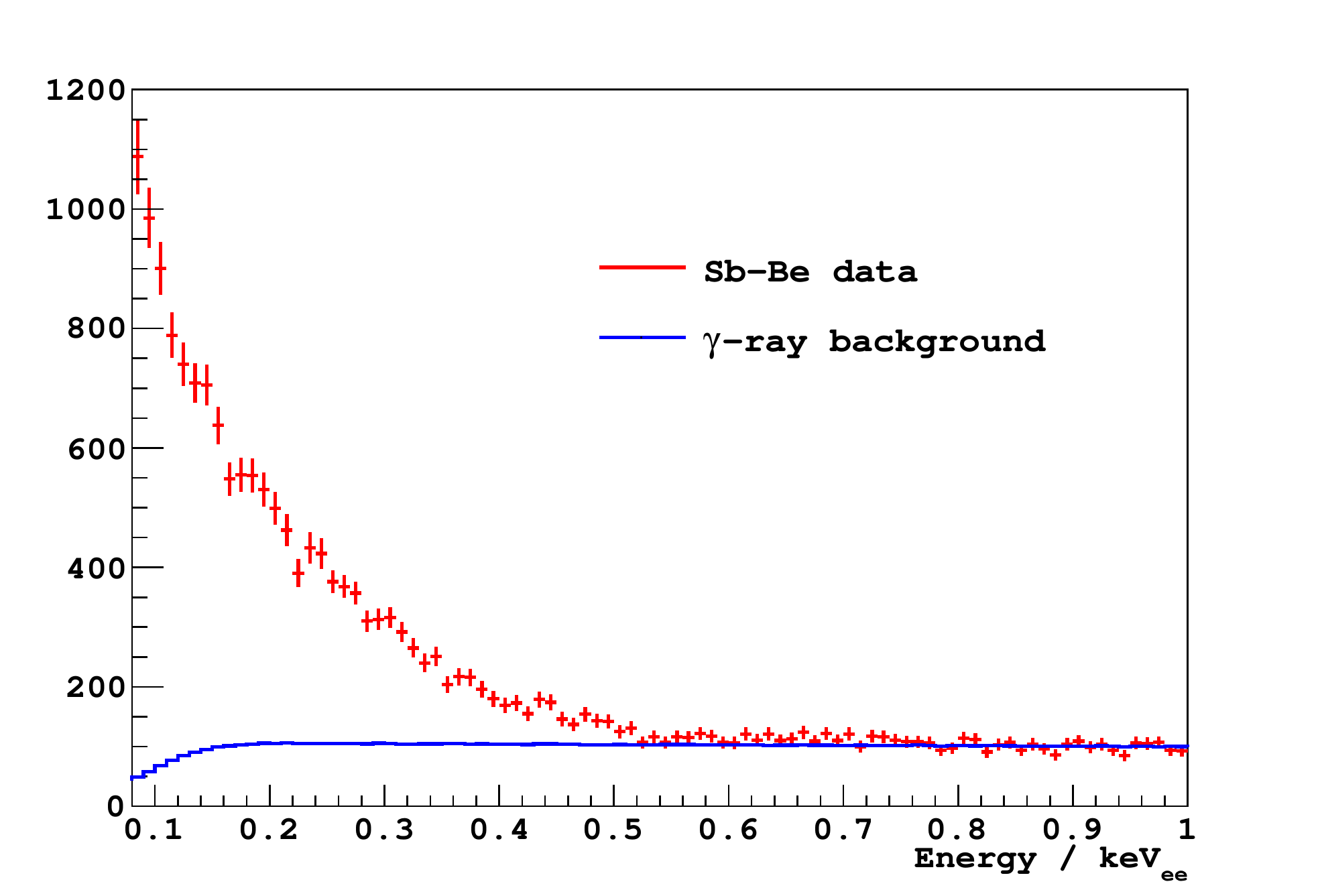}}
\caption{\small
Left: Compton scattering spectrum below 4~keV$_{ee}$ of background $\gamma$-rays from the $^{124}$Sb source. The drops at $\sim150$~eV$_{ee}$ and $\sim1.8$~keV$_{ee}$ are due to the Si atomic structure.
Right: Spectrum observed from the $^{124}$Sb source with the beryllium target (red). The blue lines shows the $\gamma$-ray background model derived from data with the aluminum target. The excess in the red spectrum is due to 24\,keV neutrons produced by the $^{9}$Be($\gamma,n$) reaction.
}
\label{fig:gamma-compton}
\end{figure}

The response of a DAMIC CCD to $\gamma$-rays and fast neutrons has been studied in a vacuum chamber at The University of Chicago. These studies have been done for the calibration of the CCD with 24\,keV neutrons produced by $\gamma$-rays from a $^{124}$Sb source impinging on a $^{9}$Be target. The configuration of the source allows to characterize the $\gamma$-ray background by replacing the beryllium with an aluminum target, which does not produce any neutrons. Figure \ref{fig:gamma-compton}(left) shows the energy spectrum from the background $\gamma$-rays. Steps are visible corresponding to the K-shell (1.8\,keV$_{ee}$) and L-shell (0.15\,keV$_{ee}$) electron binding energies. These correspond to the minimum energy deposited by $\gamma$-rays when scattering on electrons from the different atomic shells. The magnitude of the vertical drops is approximately proportional to the number of electrons in each shell. This demonstrates the capability of the detectors to resolve features in the deposited energy spectrum close to the nominal threshold. 

Preliminary studies with the $^{124}$Sb-$^9$Be neutron source have provided evidence of the CCD's ability to measure Si recoils with a maximum energy of 3.2\,keV$_r$. This can be seen in figure \ref{fig:gamma-compton}(right), where the red histogram shows the spectral excess observed with the beryllium target in place, containing the signal from neutrons in addition to the $\gamma$-ray background.

\section{Preliminary dark matter result with 2014 data}

A search for dark matter was performed using data from 36 days of exposure with two 500~$\mu$m-thick\footnote{Thinned astronomical CCD's with indium-tin oxide (ITO) coating.} (2.2~g each), and one 675~$\mu$m-thick\footnote{Un-thinned CCD without ITO coating.} (2.9~g), plus an additional 7 days of exposure with the 675~$\mu$m-thick CCD (total exposure of 0.27~kg$\cdot$day).

\begin{figure}[t]
\scalebox{0.62}{\includegraphics{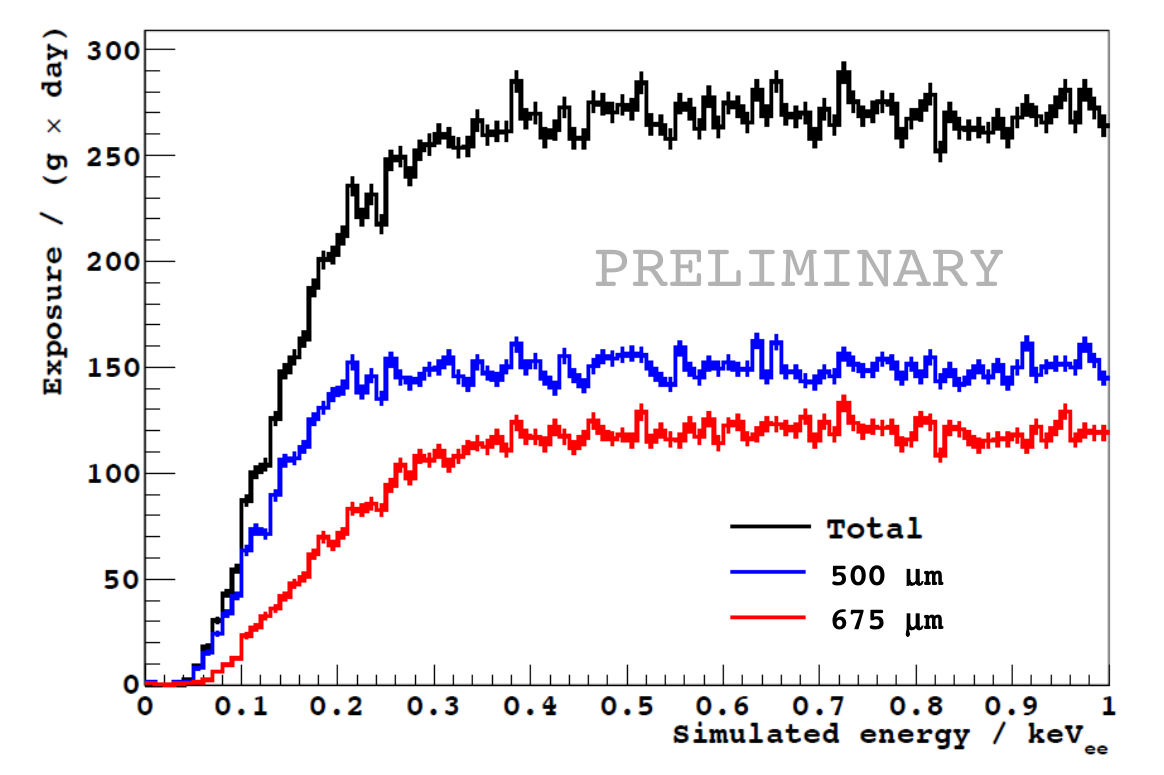}}
\scalebox{0.62}{\includegraphics{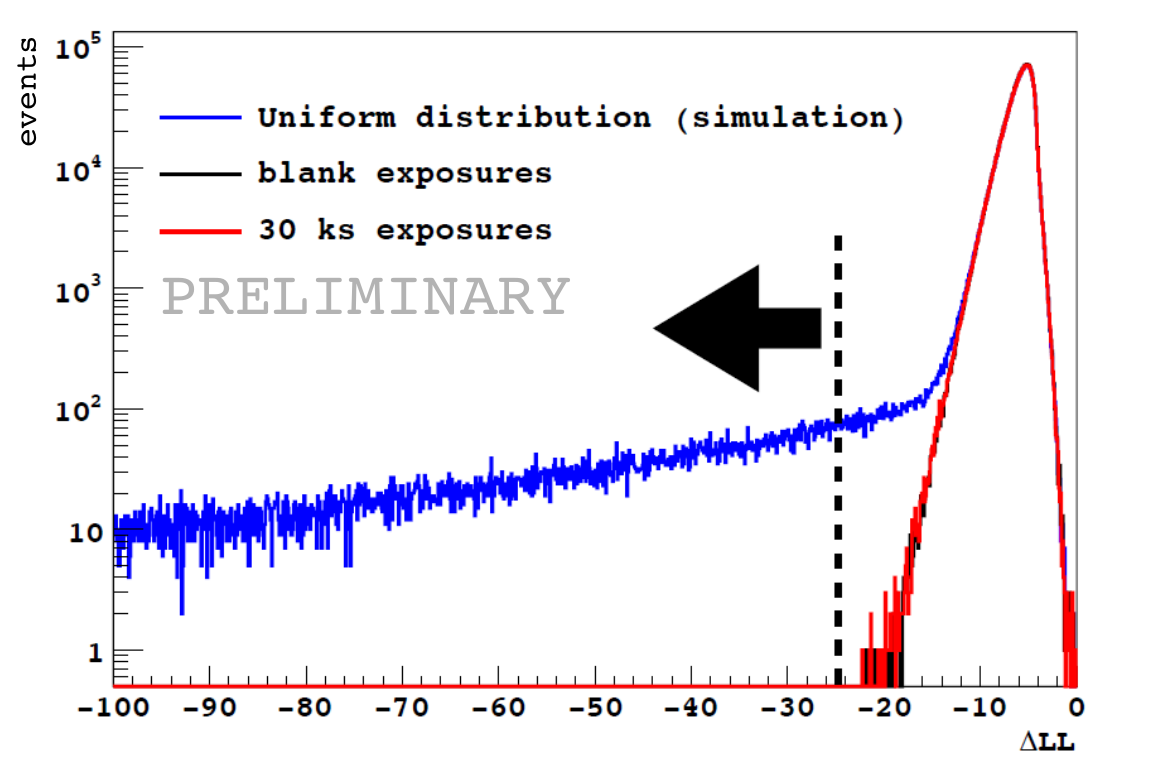}}
\caption{\small
Left: Exposures as a function of energy for the CCDs used in the dark matter search analysis. 
Right: $\Delta {\rm LL}$ distributions for events with $E<0.25$~keV$_{ee}$ in a 30~ks real exposure, blank exposures (no physical events), and uniformly distributed events simulated on the blank expesures.
}
\label{fig:exposure-DLL}
\end{figure}

WIMP candidate events are expected to deposit energies $\lesssim10$~keV and form diffusion-limited hits \cite{chavarria:2015}. These events were searched for by performing a 2D Gaussian fit to the charge distribution in a moving 7$\times$7 pixel window with the following parametrization:
$N_e\times{\tt Gauss}(x,y\;|\;\mu_x,\mu_y, \sigma)+c_{\rm ped}$.
Here, $x$ and $y$ are the pixel coordinates, and the five fit parameters are the total number of ionized electrons $N_e$ (in $e^-$), the event width $\sigma$ (in pixels) due to lateral diffusion, the mean position of the charge distribution $\mu_x$ and $\mu_y$ in each direction, and a residual local pedestal $c_{\rm ped}$. 
The difference between negative log-likelihood (LL) of the best fit, and that obtained from a fit to only a local pedestal, $\Delta {\rm LL} = {\rm LL}_{\rm bf}-{\rm LL}_{\rm ped}$, was used as discriminant.
Figure~\ref{fig:exposure-DLL}(left) shows the exposures of the CCDs used in the analysis. Figure~\ref{fig:exposure-DLL}(right) shows the $\Delta {\rm LL}$ distribution from a 30~ks exposure, and compares it to the distribution from data blanks (containing no physical events), and that from simulated events with uniform energy distributed evenly throughout the data blanks.

An unbinned likelihood fit of a standard dark matter halo model \cite{smith:1996} was performed over the data events with energies $<10$~keV$_{ee}$, letting also float a flat level background. The model assumed a local WIMP density of 0.3~GeV/cm$^{3}$, halo dispersion velocity of 220~km/s, Earth velocity of 232 km/s, and escape velocity of 554~km/s. The detector response simulation used the efficiencies extracted from the data shown in figure \ref{fig:exposure-DLL}(left), and assumed the Lindhard model \cite{ziegler:1985} to calculate nuclear recoil energies. 

\begin{figure}[t]
\scalebox{0.62}{\includegraphics{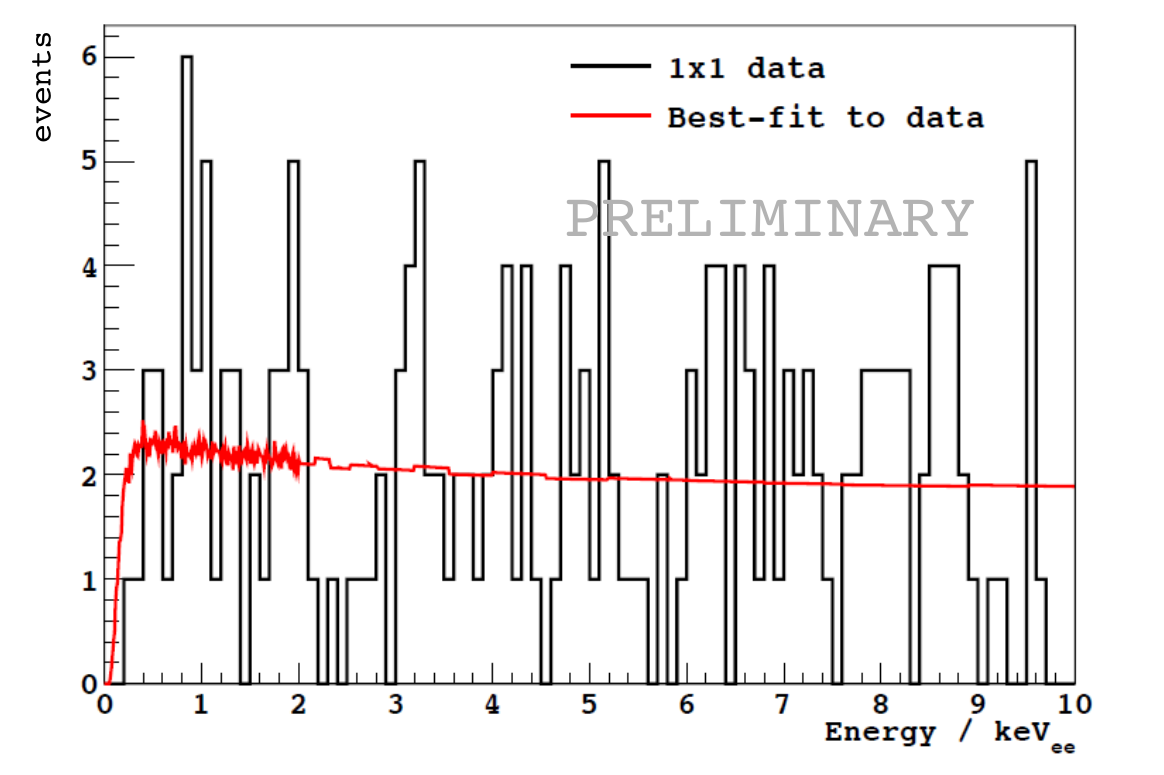}}
\scalebox{0.62}{\includegraphics{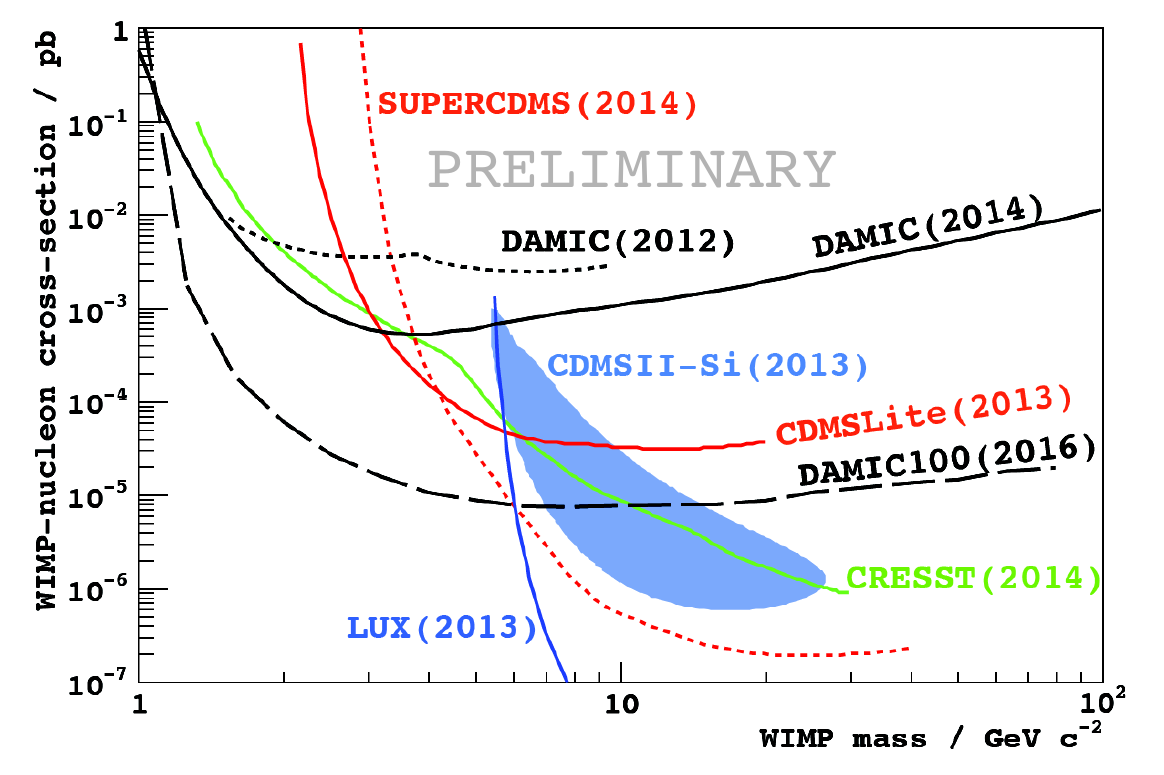}}
\caption{\small
Left: Energy spectrum of data events used in the dark matter search (black histogram) compared to the best fit energy spectrum (red line).
Right: Extracted 90\% C.L. limit (solid black line). Also shown is the projected sensitivity for DAMIC100 (black long-dash line) with a 36.5~kg$\cdot$day exposure, assuming a limiting background of $0.5$~kg$^{-1}$day$^{-1}$keV$_{ee}^{-1}$, and a threshold of 0.5~keV$_r$. Several other results are shown for comparison.
}
\label{fig:dm-results}
\end{figure}

Figure~\ref{fig:dm-results}(left) shows the energy spectrum of the data events used in the fit (black line), and the predicted best-fit spectrum (red line). The resulting best fit parameters are $M_\chi=26\pm46$~GeV/$c^2$, $\sigma_\chi=7\pm16\times10^{-4}$~pb, and $c_{\rm bg}=67\pm13$~dru, consistent with the absence of a dark matter signal. The best fit had a negative log-likelihood minimum of ${\rm -log L}_{\rm bf}=-396.5$, consistent with the value from a fit to the NULL hypothesis ($c_{\rm bg}=74\pm5$~dru, ${\rm -logL}_{\rm NULL}=-396.1$). The 90\% C.L. limit on the WIMP parameters is shown in the right-hand side plot in figure~\ref{fig:dm-results}.

\section{Expectations for DAMIC100}

The full deployment of DAMIC100 will consist of eighteen 4k$\times$4k (16~Mpix) CCDs 675~$\mu$m-thick for a total Si mass of $\sim100$~g in the current vessel and shielding at SNOLAB. The radioactive background event rate from the CCD packaging is expected to be $\ll~1$~dru leading to a limiting background  dominated by Compton scattering of external $\gamma$-rays at a predicted rate of $\sim0.5$~dru. Figure~\ref{fig:dm-results}(right) shows the expected DAMIC100 sensitivity (long-dash black line) after one year of operation assuming this level of background, and a 0.5~keV$_r$ threshold. The physics run with this new detector should begin during the first half of 2016.

\section{Conclusions}

The DAMIC collaboration has demonstrated the potential of CCDs to perform a competitive search for light WIMPs. A preliminary result from a short exposure of $\sim0.3$~kg$\cdot$day at SNOLAB has been presented. A novel method for the characterization of the CCD response to low energy $\gamma$-rays exploiting the atomic structure of silicon has been developed. Recent studies with neutron sources suggest that the detectors should be able to measure Si recoils with energies near the 50~eV$_{ee}$ threshold.

The installation of DAMIC100 is well underway. The detectors are in the process of being packaged in new low radioactive assemblies for subsequent testing. Radioactivity tests of the new CCD package have shown that the experiment will be able to achieve a limiting background rate close to $\sim$ 0.5~dru required to reach the projected sensitivity. The experiment should begin data taking during the first half of 2016.

\Acknowledgements

The DAMIC collaboration would like to thank SNOLAB and its staff for providing 
underground laboratory space and outstanding technical support, and Vale S.A. 
for hosting SNOLAB. We thank G. E. Derylo and K. R. Kuk for their contributions 
to the design, construction and installation of the detector. We are grateful 
to the following agencies and organizations for financial support: Kavli 
Institute for Cosmological Physics at the University of Chicago through grant 
NSF PHY-1125897 and an endowment from the Kavli Foundation, the Natural Sciences 
and Engineering Research Council of Canada, the Ontario Ministry of Research and 
Innovation, the Northern Ontario Heritage Fund, the Canada Foundation for 
Innovation, DGAPA-UNAM through grants PAPIIT No. IN112213 and No. IB100413, 
Consejo Nacional de Ciencia y Tecnolog\'ia (CONACYT), M\'exico, through grant 
No. 240666, the Swiss National Science Foundation through grant 153654, and the 
Brazilian agencies Coordenac\~ao de Aperfeicioamento de Pessoal de N\'ivel 
Superior (CAPES), Conselho Nacional de Desenvolvimento Cient\'ifico e 
Tecnol\'ogico (CNPq) and Fundac\~ao de Amparo \~a Pesquisa do Estado de Rio de 
Janeiro (FAPERJ). 
This work is supported by the U.S. Department of Energy, Office of Science,
Office of High Energy Physics. Fermi National Accelerator Laboratory is
operated by Fermi Research Alliance, LLC under Contract No.
De-AC02-07CH11359 with the United States Department of Energy.


\begin{thebibliography}{99}


\bibitem{Clowe:2006}
D.~Clowe {\it et al.},
{\it A direct empirical proof of the existence of dark matter},
Astr. J. Lett. 648 (2) (2006) L109. 

\bibitem{Hinshaw:2013}
G.~Hinshaw {\it et al.} [WMAP Collaboration],
{\it Nine-year wilkinson microwave anisotropy probe (WMAP) observations: Cosmological parameter results}, Astr. J. Sup. Ser. 208 (2) (2013) 19. 

\bibitem{Planck:2015}
P.A.R.~Ade, {\it et al.} [Planck Collaboration], 
{\it Planck 2015 results}. XIII. Cosmological parameters, [arXiv:1502.01589].

\bibitem{cohen:2013}
T. Cohen {\it et al.},
{\it Asymmetric Dark Matter from a GeV Hidden Sector}, Phys.Rev. D82 (2010) 056001. [arXiv:1005.1655]

\bibitem{cdmssi:2013}
R.~Agnese {\it et al.} [CDMS Collaboration],
{\it Silicon detector dark matter results from the final exposure of cdms ii}, Phys. Rev. Lett. 111 (2013) 251301. 

\bibitem{dama:2013}
R. Bernabei {\it et al.} [DAMA/Libra Collaboration],
{\it Final model independent result of dama/libra–phase1}, The Eur. Phys. J. C 73 (12) (2013) 1-11. 


\bibitem{barreto:2012}
J. Barreto {\it et al.} [DAMIC Collaboration],
{\it Direct search for low mass dark matter particles with CCDs}, Physics Letters B 711 (3-4) (2012) 264-269. 

\bibitem{chavarria:2015}
A. E. Chavarria {\it et al.} [DAMIC Collaboration],
{\it DAMIC at SNOLAB}, Physics Procedia 61, 21-33 (2015), [arXiv:1407.0347]

\bibitem{tiffenberg:2014}
J. Tiffenberg {\it et al.} [DAMIC Collaboration] "{\it DAMIC: a novel dark matter experiment"}, Proceedings, 33rd International Cosmic Ray Conference (ICRC2013), Braz.J.Phys. 44 (2014) pp.415-608


\bibitem{aguilar-arevalo:2015}
A. Aguilar-Arevalo {\it et al.} [DAMIC Collaboration], 
{\it Measurement of radioactive contamination in the high-resistivity silicon CCDs of the DAMIC experiment}, JINST 10 (2015) P08014.


\bibitem{ziegler:1985}
J. Ziegler {\it et al.},
{\it The Stopping and Range of Ions in Solids}, Stopping and Range of Ions in Matter, Vol 1, Pergamon Press, 1985. 

\bibitem{dougherty:1992}
B. L. Dougherty, 
{\it Measurements of ionization produced in silicon crystals by low-energy silicon atoms}, Phys. Rev. A 45 (1992) 2104-2107.

\bibitem{gerbier:1990}
G. Gerbier {\it et al.},
{\it Measurement of the ionization of slow silicon nuclei in silicon for the calibration of a silicon dark-matter detector}, Phys. Rev. D 42 (1990) 3211-3214. doi:10.1103/PhysRevD.42.3211. 

\bibitem{smith:1996}
J.D. Lewin, P.F. Smith, 
{\it Review of mathematics, numerical factors, and corrections for dark matter experiments based on nuiclear recoil}, Astropart. Phys. 6, 87-112 (1996).



\end{thebibliography}
\end{document}